# Osmosis, colligative properties, entropy, free energy and the chemical potential

Peter Hugo Nelson

*Physics and Biology, Benedictine University, Lisle, IL 60532*

(v.1.2 dated September 10, 2014)

## Abstract

A *diffusive* model of osmosis is presented that explains currently available experimental data. It makes predictions that distinguish it from the traditional convective *flow* model of osmosis, some of which have already been confirmed experimentally and others have yet to be tested. It also provides a simple kinetic explanation of Raoult's law and the colligative properties of dilute aqueous solutions. The diffusive model explains that when a water molecule jumps from low to high osmolarity at equilibrium, the free energy change is zero because the work done pressurizing the water molecule is balanced by the entropy of mixing. It also explains that equal chemical potentials are required for particle exchange equilibrium in analogy with the familiar requirement of equal temperatures at thermal equilibrium.

# I. Introduction

The life-science curriculum is currently under review and it has been concluded that there is a need to redesign introductory physics for the life sciences (IPLS) to better meet student needs and interests.[1] I believe that introductory physics should be the first course where life-science majors are introduced to quantitative scientific modeling, but the traditional introductory physics curriculum was not designed for them. Biological examples and applications have been added over the years, but most of them are not recognized as being relevant by biologists.[2] In a recent survey conducted by the Association of American Medical Colleges, "transport processes" (diffusion, osmosis, etc.) was identified as being the second most important topic overall after "nucleic acids".[3] However, this physics topic is usually absent from current IPLS courses.[4]

Thermodynamics is also an important topic for life-science majors, but traditional textbook presentations are not seen as being productive in an authentic biological or chemical context.[5] While temperature is a concept that seems intuitive to most students, the meaning of chemical potential is elusive.[6] Thermal conduction (heat transfer) is proportional to the temperature gradient, and mass transport (particle transfer) is proportional to the chemical potential gradient. The equivalence of these two concepts should be presented to students in a straightforward manner. Randomness,[7] entropy, free energy and the chemical potential are key thermodynamic concepts that should be integrated into the IPLS curriculum.

The "marble game" is a kinetic Monte Carlo simulation that provides a new pathway to quantitative scientific modeling that can be used from the very first class.[8] It provides an introductory model that can be used to build a computational and mathematical framework that spans the science, technology, engineering and math (STEM) disciplines. It was recently successfully tested, by asking students to *derive* a novel theory of osmosis under final exam conditions.[8] In this paper, that model of osmosis is simplified for use in IPLS courses in a manner that also introduces students to basic thermodynamic concepts including: how energy differences affect the rates of molecular processes; entropy; free energy; and the chemical potential.

The diffusive model of osmosis presented here also provides an opportunity to teach science like we do science. This diffusive model of osmosis[9] is currently controversial, providing a conceptual picture of osmosis that conflicts[10,11] with the traditional biophysics[12] and physics[13] textbook descriptions. It is hoped that addressing this controversy in teaching materials will inspire life-science students to further pursue quantitative scientific modeling.[9]

The remainder of this paper is organized as follows. Section II introduces the diffusive model of osmosis and an equation is derived for the osmotic swelling/shrinking of red blood cells (RBCs) within the constant-pressure Gibbs ensemble. The "marble gravity game" is introduced in Sec. III to show how a mechanical energy difference can affect the jump rate of marbles between two boxes using the concept of an energy factor. In Sec. IV, the energy factor is used to show how a pressure difference affects osmosis within the constant-volume Helmholtz ensemble, and the van't Hoff equation for the equilibrium pressure difference (osmotic pressure) is derived from the diffusive model. A key simplifying concept in the preliminary presentation (Sec. II) is the concept of an "effective water concentration" that is related to osmolarity via a deceptively simple equation (3). In Sec. V a kinetic explanation of that equation is presented. This conceptual framework leads to a derivation of a more accurate "Raoult's law version" of the van't Hoff equation, which in turn provides a kinetic explanation of the entropy of mixing, Gibbs free energy and the chemical potential of ideal solutions. Kinetic models of liquid-vapor and solid-liquid coexistence are presented in Sec. VI, and Raoult's law is derived within the Gibbs ensemble, providing a consistent set of kinetic models for all of the colligative properties. In Sec. VII the diffusive (marble game) model of osmosis is compared with the traditional hydrodynamic flow model, identifying testable hypotheses for distinguishing between them experimentally and computationally. Section VIII presents the predictions of the diffusive model of osmosis for tracer counter permeation (TCP) experiments. Finally, Sec. IX discusses the relationship between the *diffusive* and traditional hydrodynamic *flow* models of osmosis and offers some concluding remarks.

## II. Osmotic swelling/shrinking – water diffusion

### A. A diffusive model of osmosis

Currently, there are two competing explanations of osmosis. One describes osmosis as a *diffusive* process and the other models osmosis as a convective *flow*. Introductory college-level chemistry and physiology textbooks[14] typically describe osmosis as the *diffusion* of water from high to low water concentration (low to high osmolarity). However, those explanations do not provide a quantitative model.[10,11] The (current) consensus view of the biophysics[12] and physics[13] communities is that osmosis is a pressure-driven convective *flow* of water through a narrow water-selective pore.[10,11]

The marble game model of osmosis (mentioned in the introduction) is based on the opposite assumption – that osmosis can indeed be modeled by a kinetic description of water *diffusion*.[8,9,15] Convective laminar *flow* is predicted by the Hagen–Poiseuille equation for the pressure-driven flow of a fluid, such as the water in a pipe, or the blood in an artery or vein. *Diffusion* is predicted by Fick's law of diffusion and is caused by random thermally activated "jumps".[8,15]

Water is the most important molecule for life as we know it and osmosis is the selective transport of water. In physiology, osmosis primarily occurs by permeation of water through the pores of aquaporin proteins (Fig. 1) that have a selectivity filter that is narrow enough to allow only a single file of water molecules to pass through.

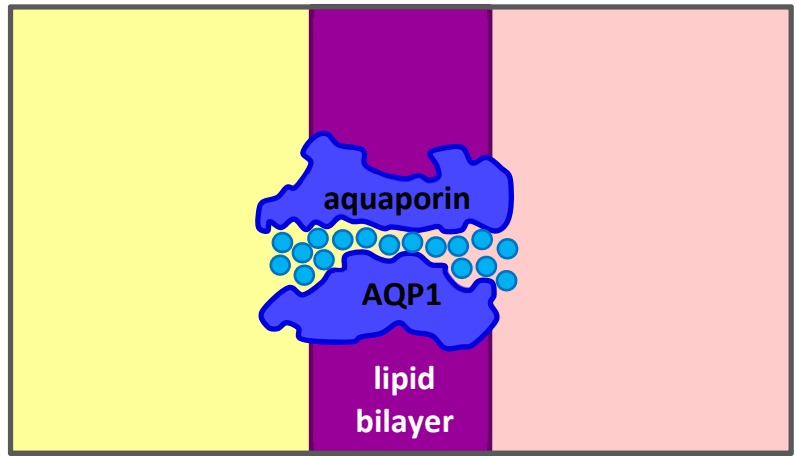

Fig. 1. Schematic diagram of an AQP1 aquaporin protein (water channel) imbedded in a lipid bilayer membrane separating two solutions with differing (effective) water concentrations (after Murata *et al.*[16]). The aquaporin provides a single-file pathway (shown in cross-section) for water molecules (circles) that makes the membrane permeable only to water (semipermeable).

Permeation through the selectivity filter can be summarized by a knock-on jump mechanism (Fig. 2), making it a physical situation that can be modeled by the marble game.[8] $k$ is the knock-on jump rate constant and $c_{w_1}$ and $c_{w_2}$ are the *effective* water concentrations in box 1 and box 2 respectively. **Please note**, throughout this paper, the numerical subscripts indicate box number (see Fig. 3).

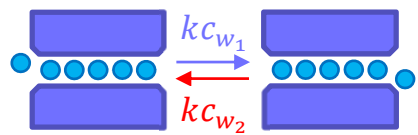

Fig. 2. Schematic diagram of an AQP1 aquaporin selectivity filter showing the knock-on jump mechanism for water permeation. The net effect is that a water molecule reversibly jumps through the selectivity filter.

Although originally conceived as a single elementary step,[17,18] the transitions shown in Fig. 2 need not be elementary. Just like the jumps in the original marble game (see Figure 1.4 of Module 1),[15] the transition shown in Fig. 2 can be the result of many smaller translocations. At timescales longer than a single knock-on jump transition, this complex single-file process can be summarized as Fickian diffusion.[19]

## B. Finite difference (FD) model of a red blood cell in solution

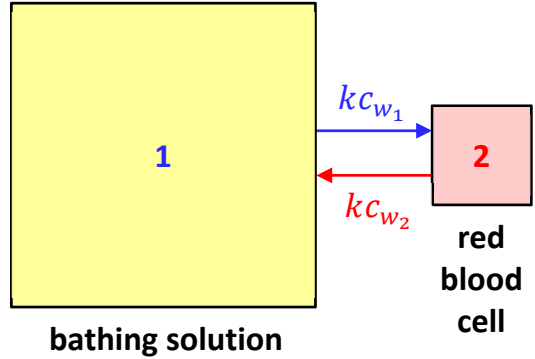

**Fig. 3. Finite difference (FD) diagram of a red blood cell (box 2) floating in a large bathing solution (box 1). The water in the red blood cell has an effective concentration $c_{w_2}$ and the bathing solution has a constant effective water concentration $c_{w_1}$.**

Osmotic swelling/shrinking of a RBC will be modeled in the Gibbs ensemble (constant $(T, P)$)[20] as the two boxes are at the same pressure, so that the pressure difference between boxes is $\Delta p = p_2 - p_1 = 0$. By inspecting Fig. 3, we can write[21]

$$\delta c_{w_2} = k(c_{w_1} - c_{w_2})\delta t \qquad (1)$$

or in differential form

$$\frac{dc_{w_2}}{dt} = -k\Delta c_w \qquad (2)$$

where $\Delta c_w = c_{w_2} - c_{w_1}$. Equation (2) is a form of Fick's first law of diffusion.[15]

The effective water concentration $c_{w_2}$ can be related to the osmolarity (or osmotic concentration) of solute particles $c_{s_2}$ in box 2 by

$$c_{w_2} = c_w^* - c_{s_2} \qquad (3)$$

where $c_w^*$ is the concentration of pure water. Similarly, $c_{w_1} = c_w^* - c_{s_1}$, and $\Delta c_s = c_{s_2} - c_{s_1}$. Hence, Eq. (2) can be written as

$$\frac{dc_{w_2}}{dt} = k\Delta c_s \qquad (4)$$

When a water molecule jumps into the RBC, the RBC's volume $V_2$ changes by $\delta V_2 = v_w = 1/(N_A c_w^*)$, the volume of a liquid water molecule $v_w$, where $N_A$ is Avogadro's number. It is also assumed that the membrane is impermeable to all solutes on the timescale of the experiment, so that the number of solute particles in the RBC is constant, so that $c_{s_2} = c_{s_{2_0}}/V_r$, where the relative volume of the RBC is defined as $V_r = V_2/V_{2_0}$ and $V_{2_0}$ is the initial volume of the RBC.

Hence, the FD water permeability equation for the small change in relative volume during a short time $\delta t$ is

$$\delta V_r = \frac{k}{c_w^*}\left(\frac{c_{s_{2_0}}}{V_r} - c_{s_1}\right)\delta t \qquad (5)$$

or in differential form

$$\frac{dV_r}{dt} = \mathcal{P}_f \frac{A_2}{V_{2_0}} \bar{V}_w \left(\frac{c_{s_{2_0}}}{V_r} - c_{s_1}\right) \qquad (6)$$

where

$$\mathcal{P}_f = \frac{kV_{2_0}}{A_2} \qquad (7)$$

is the filtration permeability (or osmotic permeability) of the RBC membrane.

Equation (6) is the fifth equation (unnumbered) in the left-hand column of page 1310 of Mathai et al.[22] $A_2/V_{2_0}$ is the initial surface area $A_2$ to volume $V_{2_0}$ ratio of the RBC. $\bar{V}_w = 1/c_w^*$ is the molar volume of pure water. Mathai et al.[22] successfully fitted Eq. (6) to RBC experimental data, finding the osmotic water permeability of RBCs to be $\mathcal{P}_f = 22.8 \times 10^{-3}$ cm/s, giving a jump rate constant of $k = 500$ s$^{-1}$ for the two-box system of Fig. 3. Thus, the predictions of the diffusive model of osmosis within the Gibbs ensemble ($\Delta p = 0$) are consistent with the traditional hydrodynamic flow model and they have already been confirmed experimentally in Nobel Prize winning research.[22]

From a pedagogical perspective, it is important to note that equation (5) predicts that osmotic equilibrium within the Gibbs ensemble corresponds to

$$V_r^{eq} = \frac{c_{s_{2_0}}}{c_{s_1}} \qquad (8)$$

I.e. the relative osmotic swelling/shrinking is determined by the initial osmolarity ratio. It should also be noted that Eq. (8) is only valid for values of $V_r^{eq}$ that are physically possible, e.g. if $V_r^{eq}$ is too large, the cell will burst (lyse).

The diffusive flux within the Gibbs ensemble can thus be written in the same form as Fick's first law of diffusion[15]

$$j = -\mathcal{P}_f \Delta c_w = \mathcal{P}_f \Delta c_s \qquad (9)$$

or in terms of the volumetric flux

$$Q = L_p \Delta \pi \qquad (10)$$

where the hydraulic permeability $L_p$ and the osmotic pressure difference $\Delta \pi$ will be defined in Sec. IV below.

The Gibbs ensemble ($\Delta p = 0$) relates to a situation where the two boxes have the same hydrostatic pressure. An everyday example where that assumption is not true is the crisping of a limp celery stalk that is freshly cut and placed in a glass of tap water. As the limp celery stalk becomes firm, a turgor pressure develops that eventually stops the net osmotic diffusion of water into its cells. This phenomenon can be understood in terms of the energetics of the jumps of water molecules between the boxes. Students can be introduced to this concept with the marble gravity game.

## III. Marble gravity game

The marble gravity game provides an intuitive introduction into how energy differences can affect the jump rates between boxes.[9] The two boxes are at different heights so that uphill jumps require additional energy (Fig. 4). The isothermal atmosphere is the textbook example[23] that corresponds to the gravity marble game. The jumps between boxes are caused by simple Brownian motion (molecular diffusion). Figure 5 shows an FD diagram for this system.

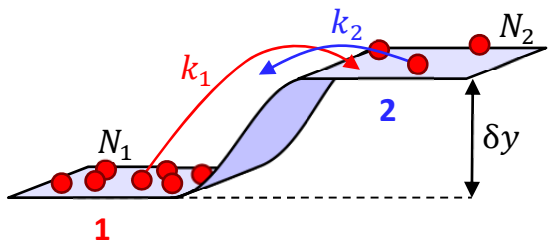

**Fig. 4. Schematic representation of the marble gravity game. The jump rate constant in the uphill direction (from box 1 → 2) is $k_1$ and $k_2$ is the jump rate constant in the downhill direction (from box 2 → 1). The marbles each have mass $m$ and the two boxes are separated by height $\delta y$.**

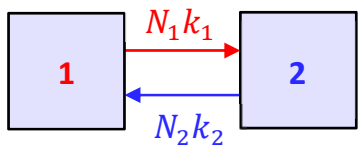

**Fig. 5. FD diagram of the marble gravity game.**

$N_1$ and $N_2$ are the number of marbles of mass $m$ in boxes 1 and 2 respectively and $k_1$ and $k_2$ are the jump rate constants for jumps originating in boxes 1 and 2 respectively that are separated by a small height difference $\delta y$. By inspecting FD diagram 5,[21] the change in the number of molecules (marbles) in box 2 during a short time $\delta t$ is given by

$$\delta N_2 = (N_1 k_1 - N_2 k_2)\delta t \tag{11}$$

At equilibrium, the number of molecules in each box is constant in time and

$$\frac{N_1}{N_2} = \frac{k_2}{k_1} \quad (12)$$

For the isothermal atmosphere, the pressure difference between the two boxes is

$$\delta p = p_2 - p_1 = -\rho g \delta y \quad (13)$$

so that $\delta N = N_2 - N_1$ is given by

$$\delta N = -N\delta\psi \quad (14)$$

where the small dimensionless energy step $\delta\psi$ is given by

$$\delta\psi = \frac{\delta E}{k_B T} = \frac{mg\delta y}{k_B T} \quad (15)$$

where $\delta E = mg\delta y$ is the potential energy difference between the two boxes caused by a gravitational field of strength $g$. $k_B T$ is the thermal energy (Boltzmann's constant times absolute temperature).

The ratio of rates defines the energy factor $\varepsilon$, which is

$$\varepsilon = \frac{k_2}{k_1} \quad (16)$$

where

$$\varepsilon = 1 - \delta\psi \quad (17)$$

Without loss of generality, we can let $k_2 = k$, so that $k_1 = \varepsilon k$ and the jumps between boxes separated by a small energy difference $\delta E$ can be summarized by Fig. 6.

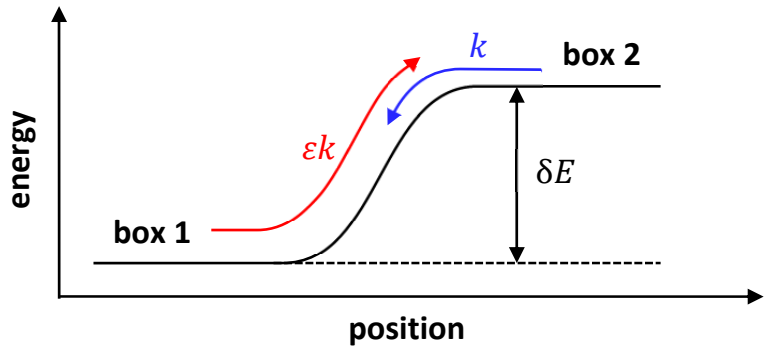

**Fig. 6. Simplified schematic energy diagram of the marble gravity game. The diagram shows a situation where marbles in box 2 have an energy $\delta E$ higher than box 1.**

The energy diagram shown in Fig. 6 applies to any situation where there is a small energy difference $\delta E$ between the boxes. As described by Eqs. (16) and (17), the uphill jump rate constant is reduced by an energy factor $\varepsilon$ that depends on the small energy step $\delta E$. $\delta \psi$ must be small compared with 1 for equation (17) to be valid. If the dimensionless energy difference is not small then Eq. (14) can be integrated to give

$$\frac{N_2}{N_1} = e^{-\Delta \psi} \tag{18}$$

where $\Delta \psi = \psi_2 - \psi_1$ is the dimensionless energy difference, which need not be small. In this more general case, the energy factor $\varepsilon = e^{-\Delta \psi}$ becomes the familiar Boltzmann factor. For a small dimensionless energy step $\Delta \psi \to \delta \psi$ and

$$\varepsilon = e^{-\delta \psi} \approx 1 - \delta \psi \tag{19}$$

so that the energy factor $\varepsilon$ is a linearized Boltzmann factor for small dimensionless energy steps $\delta \psi$, such as those encountered in osmosis.

## IV. Osmotic pressurization

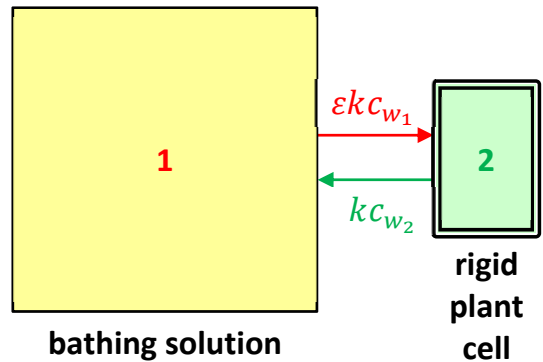

**Fig. 7.** FD diagram of a rigid plant cell (box 2) in contact with a large bathing solution (box 1). The water in the cell has an effective water concentration $c_{w_2}$ and the bathing solution has an effective water concentration $c_{w_1}$. There may also be a hydrostatic pressure difference $\Delta p = p_2 - p_1$ between the boxes that determines the value of the energy factor $\varepsilon$.

When a limp celery stalk is cut and placed in a glass of tap water, the crisping that occurs can be modeled within the Helmholtz ensemble (constant $(T, V)$),[24] if we assume that the plant cells are perfectly rigid. Uphill jumps of water molecules now require $pV$ work of magnitude

$$\delta E = v_w \Delta p \tag{20}$$

where $v_w$ is once again the volume of a single water molecule and $\Delta p = p_2 - p_1$ is the pressure difference between the rigid plant cell (box 2) and the bathing solution (box 1)

Equations (15), (17) and (20) can be combined with Fig. 7 to provide a diffusive model of osmosis to predict that the osmotic permeation rate is

$$\frac{dc_{w_2}}{dt} = k(\varepsilon c_{w_1} - c_{w_2}) \tag{21}$$

or that the molar flux is

$$j = -\mathcal{P}_f \left( \frac{\Delta p}{RT} - \Delta c_s \right) \tag{22}$$

when $c_{w_1} \cong c_w^*$, so that at equilibrium

$$\Delta p = RT \Delta c_s \tag{23}$$

The equilibrium pressure difference between box 2 and pure water is *defined* as the osmotic pressure (the van't Hoff equation)[12,13]

$$\pi = c_s RT \tag{24}$$

where we have dropped the subscript 2 for the solute concentration in box 2. The osmotic pressure $\pi$ indicates the chemical potential of water in an analogous manner to how oxygen tension (partial pressure) $p_{O_2}$ indicates the chemical potential of oxygen dissolved in plasma.[21]

Hence, the volumetric flux is given by Starling's law of filtration

$$Q = -L_p(\Delta p - \Delta \pi) \tag{25}$$

where

$$L_p = \frac{\bar{V}_w A_2}{RT} \mathcal{P}_f = \frac{\bar{V}_w V_{2_0}}{RT} k \tag{26}$$

is the hydraulic permeability of the membrane.

In general the osmotic pressure difference is given by

$$\Delta \pi = \pi_2 - \pi_1 \tag{27}$$

when box 1 has a nonzero osmolarity. Equation (25) shows the formal equivalence of the hydrostatic pressure difference $\Delta p$ and the osmotic pressure difference $\Delta \pi$, so that each can drive osmotic diffusion into/out of box 2 with the same jump rate constant $k$ (or permeability $\mathcal{P}_f$ or $L_p$) as for the original diffusive model (Eq. (1) etc.) and equilibrium is reached when $\Delta p = \Delta \pi$. However, the osmotic pressure difference $\Delta \pi$ is not a real pressure difference, it is simply a thermodynamic measure of the osmolarity difference $\Delta c_s$ (or the effective water concentration difference $\Delta c_w$). The relationship between them is summarized by an alternate form of the van't Hoff equation.

$$\Delta \pi = \Delta c_s RT \tag{28}$$

## V. Solute blocking and effective water concentration

### A. Raoult's law and colligative properties

If we consider diluting one liter of pure $H_2O$ with 0.3 moles of HDO (semiheavy water), glucose or sucrose, the volumetric dilution effect for glucose is about six times that for HDO and sucrose has about twelve times the effect of HDO. However, as Raoult discovered, all solutes have the same small effect on the vapor pressure of $H_2O$ as semiheavy water (HDO), even though they can be many hundreds (or even thousands) of times larger than $H_2O$. Raoult's law says that somehow their size doesn't matter. What actually counts is their mole fraction in solution. A practical consequence is that if you have a mystery powder (a pure compound), you can use Raoult's law to find its molecular weight. All you have to do is weigh out a small sample to get its mass in grams. You then dissolve it in 1 L of pure water and measure the ($H_2O$) vapor pressure. Raoult's law (58) (derived below) then tells you how many moles of the substance were dissolved. Raoult's law is thus conceptually different from Henry's law which says that the vapor pressure of a solute (e.g. of oxygen) is proportional to its concentration.[21]

The vapor pressure depression predicted by Raoult's law is one of the colligative properties that depend only on the mole fraction of solute particles in solution (and not their size or their chemical identity). These colligative properties include

1. Vapor pressure depression
2. Boiling point elevation
3. Freezing point depression
4. Osmotic pressure

According to the van't Hoff equation (24), osmotic pressure $\pi$ is directly proportional to the osmotic concentration or osmolarity $c_s$ of solute particles and not on any of the solute particles' physical or chemical properties. A tiny electrically charged sodium ion ($Na^+$) counts that same as a large hemoglobin (Hb) molecule. This is quite remarkable because $Na^+$ is a small ion that actually has a negative partial molar volume because it makes the open structure of liquid water collapse around it and a hemoglobin molecule has a partial molar volume 670 *times* the size of a water molecule! Similarly, the other colligative properties in the list depend only on the osmotic concentration and not on the properties of the solute so long as the solution is dilute. The osmotic concentration is a real concentration, number of solute particles $n_s$ per volume $V$

$$c_s = \frac{n_s}{V} \tag{29}$$

whereas the effective water concentration defined by Eq. (3)

$$c_w = c_w^* - c_s \tag{30}$$

is not – except under the special circumstance that the solute has the same partial molar volume as water (e.g. HDO). This phenomenon can be explained using the solute-blocking model of osmosis as summarized pictorially in Fig. 8.

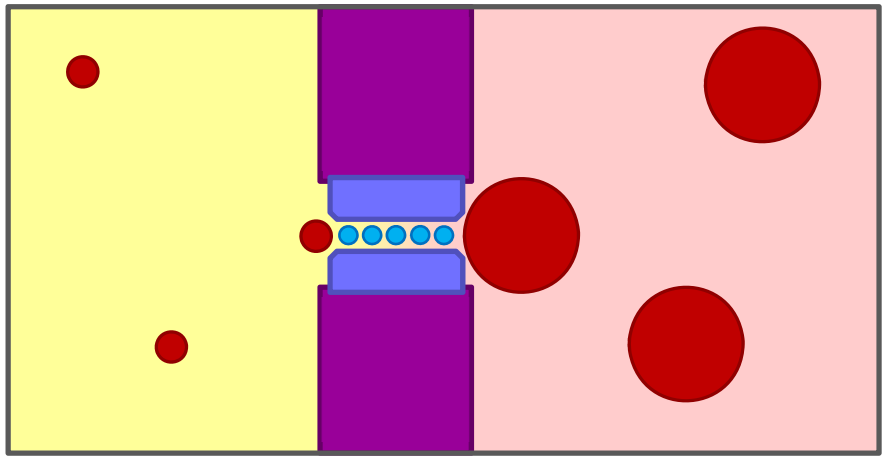

Fig. 8. Schematic diagram of an AQP1 aquaporin selectivity filter showing both ends being temporarily blocked by solute molecules. The two solutions have the same mole fraction of solutes. Water molecules fill the remainder of the solutions, but they are only shown in the selectivity filter. The size and chemical composition of the solute particles (larger red/darker circles) does not matter. The only thing that matters is how likely they are to be in position to block diffusion of water molecules though the pore (as shown). When the solute particles are away from the pore entrances pure water can diffuse through the pore by concerted jumps at the same rate as for pure water.

## B. Solute blocking and mole fraction

The reason that dissolved solute molecules slow down the knock-on jumps of water molecules through aquaporins can be understood by considering Figs. 9 and 10.

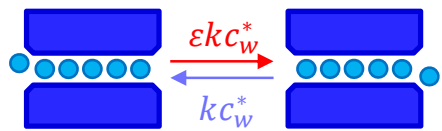

Fig. 9. Schematic diagram of an AQP1 aquaporin selectivity filter showing the knock-on jump mechanism for pure water permeation (based on Fig. 1).

Figure 9 is identical to Fig. 1 except for the fact that both effective water concentrations ($c_{w_1}$ and $c_{w_2}$) have been replaced with the concentration of pure water $c_w^*$ and the energy factor $\varepsilon$ has been added to account for the pressure difference $\Delta p$. The idea is that if both boxes contain pure water, then a water molecule is always the nearest molecule to the pore openings (i.e. the water molecule shown to the left of the selectivity filter on the left-hand side of Fig. 9 and the water molecule shown to the right of the selectivity filter on the right-hand side of Fig. 9). If the solutions contain solutes, then there is chance that there will be a solute molecule blocking the entrance to the pore as shown in Fig. 10.

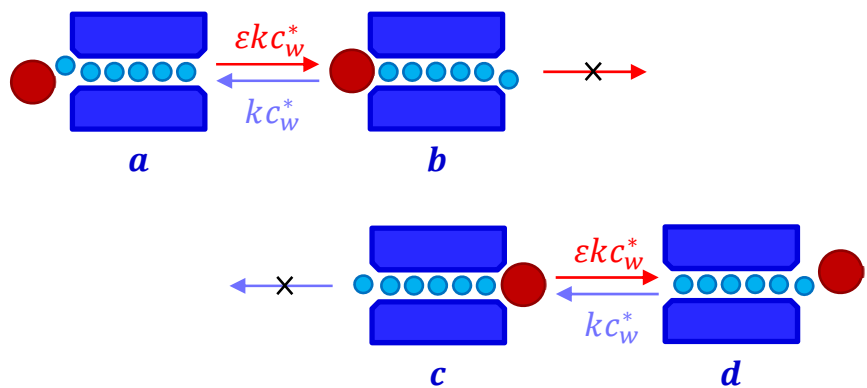

**Fig. 10. Schematic diagram of an AQP1 aquaporin selectivity filter showing how the knock-on jump mechanism for water permeation can be blocked by solute molecules. A solute in box 1 can block jumps from box 1 → 2 (top row), whereas a solute in box 2 can block jumps from box 2 → 1 (bottom row), as indicated by the crossed-out arrows.**

Figure 10 shows the reversible transitions that are possible when one end of the pore becomes blocked by a solute molecule. As shown in the top row, jumps from box 2 → 1 (states $b \to a$) are possible even if the pore entrance on the box 1 side is blocked. The reverse transition $a \to b$ (a jump from box 1 → 2) is also possible, but once the selectivity filter is blocked on the box 1 side (state $b$), further jumps of water molecules from box 1 → 2 are not possible as indicated by the crossed out left-to-right red arrow in the top row.

The second row in Fig. 10 shows that jumps from box 1 → 2 (states $c \to d$) are possible even if the pore entrance on the box 2 side is blocked. The reverse transition $d \to c$ (a jump from box 2 → 1) is also possible, but once the selectivity filter is blocked on the box 2 side (state $c$), further jumps of water molecules from box 2 → 1 are not possible as indicated by the crossed out right-to-left blue arrow in the bottom row.

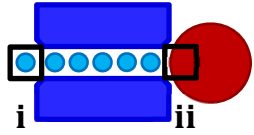

Fig. 11. Schematic diagram of an AQP1 aquaporin selectivity filter separating two nanoscopic boxes (i and ii) that are the volume $v_w$ of a single water molecule. The box on the left (box i) is shown occupied by a water molecule and the box on the right (box ii) is filled with a portion of a larger solute molecule.

In the marble game model of osmosis, the water molecules jump between two boxes. Those boxes can be any size. Fig. 11 shows what happens if we make the boxes the size of a single water molecule with volume $v_w$. When the boxes are that size, they are either full of pure water (one water molecule on average) or they contain a portion of a solute molecule. As shown in Fig. 11, box i contains pure water (concentration $c_i = c_w^*$) and box ii contains no water $c_{ii} = 0$ and jumps from box ii → i are blocked.

For the system shown in Fig. 11, there are basically only two possibilities, either there is a solute blocking the pore entrance from box ii, or there is pure water next to it. In the first case, the aquaporin is blocked (but only for jumps in the ii → i direction), and in the other case, permeation can proceed from box ii → i with box ii being full of pure water (Fig. 9). If we want to find the average (unidirectional) jump rate from box ii → i, we need the probability that a solute molecule is occupying the water-sized box ii (as shown in Fig. 11). If we assume that the solute molecules interact with the aquaporin entrance and with water molecules in a similar manner to water molecules (an ideal solution), then the probability of any one of the solute molecules in a macroscopic box 2 occupying nanoscopic box ii will be the same as the probability of any one of the water molecules occupying nanoscopic box ii. If those are the only two choices, then the probability of a solute molecule occupying box ii will be given by its mole fraction

$$x_s = \frac{n_s}{n_w + n_s} \tag{31}$$

where $n_s$ is the number of solute particles in the macroscopic box 2 and $n_w$ is the number of water molecules in box 2, where we have dropped the subscript for box 2 and box 1 contains pure water. The mole fraction of water in macroscopic box 2 is

$$x_w = \frac{n_w}{n_w + n_s} = 1 - x_s \tag{32}$$

which is the probability that nanoscopic box ii contains pure water. $x_s$ is the probability that box ii does not contain pure water.

## C. Solute-blocking model of osmosis

Figure 12 shows an FD diagram of a rigid plant cell (box 2) in contact with a bath containing pure water. The jump rate from box $1 \to 2$ is the same as the original marble game, but the jump rate from box $2 \to 1$ is now indicated using the mole fraction $x_w$ of water in box 2 to account for the fraction of jumps that are blocked by the presence of the solvent.

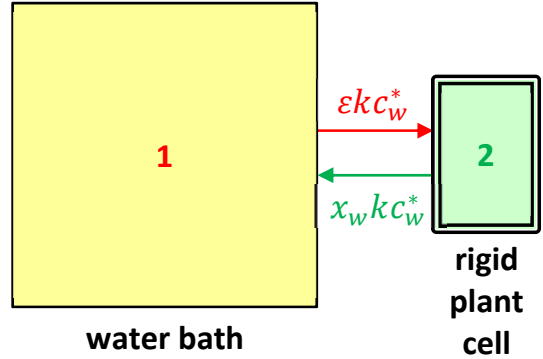

Fig. 12. FD diagram of the solute-blocking model of a rigid plant cell (box 2) in contact with a bath of pure water (box 1). The water in the cell has a mole fraction $x_w$, which reduces the jump rate from box $2 \to 1$ by a factor $x_w$ compared with pure water. There may also be a hydrostatic pressure difference $\Delta p = p_2 - p_1$ between the boxes that reduces the jump rate from box $1 \to 2$ by an energy factor $\varepsilon$.

By inspecting Fig. 12, the condition for equilibrium is

$$\varepsilon = x_w \tag{33}$$

Equation (33) provides a particularly simple (and important) explanation of the origin of osmotic pressure. It indicates that two fractions are equal at equilibrium. $\varepsilon$ is the fraction of all the molecules in box 1 that have enough energy to overcome the energy difference $\delta E = v_w \Delta p$ and $x_w$ is the fraction of all the molecules in box 2 that are water.

Substituting Eqs. (15), (17), (19), (20) and (32) into equation (33) we obtain

$$\Delta p = x_s c_w^* RT \tag{34}$$

at equilibrium, which the "Raoult's law version" of the van't Hoff equation (23).

## D. Effective water concentration and thermodynamics

For dilute solutions, $n_s + n_w \approx n_w^*$, where $n_w^*$ is the number of moles of pure water in volume $V$. Using the definition (31) of $x_s$ and the definition of concentration (that $c_w^* = n_w^*/V$ and $c_s = n_s/V$) we find that

$$x_s c_w^* \approx c_s \tag{35}$$

and the more accurate Eq. (34) reduces to the van't Hoff equation (24) for dilute solutions.

Substituting $x_s = 1 - x_w$ into Eq. (35) we find that

$$x_w c_w^* \approx c_w^* - c_s = c_w \qquad (36)$$

which is the effective water concentration in box 2, showing that the solute-blocking model of osmosis is equivalent to the diffusive model of osmosis in Sec. II.

As noted above, equation (33) is a particularly simple relationship describing osmotic equilibrium. It can be made more accurate by relaxing the assumption that the dimensionless energy step is small. In that case, the energy factor becomes the Boltzmann factor $\varepsilon = e^{-\Delta\psi}$ and Eq. (33) becomes

$$x_w = e^{-\Delta\psi} \qquad (37)$$

Substituting in the definition of the dimensionless energy step $\Delta\psi = \bar{V}_w \Delta p / RT$ and rearranging we obtain

$$\bar{V}_w \Delta p + RT \ln x_w = 0 \qquad (38)$$

Equation (38) is an alternate explanation of osmotic equilibrium in terms of the equality of two energies that cancel at equilibrium. The first term in equation (38) is the mechanical work done (per mole)

$$\Delta W = \bar{V}_w \Delta p \qquad (39)$$

moving water from box 1 → 2 through a pressure difference $\Delta p$. The second term in equation (38) is the free energy decrease when the water is "diluted" in box 2.

$$-T\Delta S_{\text{mix}} = RT \ln x_w \qquad (40)$$

where

$$\Delta S_{\text{mix}} = -R \ln x_w \qquad (41)$$

is the entropy of mixing (per mole of water molecules). The free energy change $\Delta F$ going from box 1 → 2 within the Helmholtz ensemble is thus

$$\Delta F = \Delta W - T\Delta S_{\text{mix}} \qquad (42)$$

At equilibrium, this Helmholtz free energy change is zero (Eq. (38)), so that the work done pressurizing the water is balanced by the entropy of mixing. Within the solute-blocking model of osmosis, thermodynamic equation (42) is a direct consequence of kinetic equilibrium in the model system of Fig. 12 that is summarized by equation (33).

Another important way of describing osmotic equilibrium, is that the chemical potential of water[25]

$$\mu_w = \left(\frac{\partial F}{\partial n_w}\right)_{V,T} \tag{43}$$

is the same in both boxes, where $n_w$, $V$, and $T$ are respectively, the number of moles of water, the volume and temperature of each box. The equality of the chemical potentials is a consequence of the Helmholtz free energy having a minimum with respect to particle exchange at equilibrium within the Helmholtz ensemble as $\Delta\mu_w = \mu_{w_2} - \mu_{w_1} = 0$ at equilibrium (also see below).

### E. Ideal solution thermodynamics

In summary, when the two boxes are in thermal equilibrium the temperatures are the same

$$T_2 = T_1 \tag{44}$$

and when they are in particle exchange equilibrium the chemical potentials are the same

$$\mu_2 = \mu_1 \tag{45}$$

Box 1 is pure water, hence

$$\mu_1 = \mu_w^* \tag{46}$$

where $\mu_w^*$ is the chemical potential of the pure water reference state and from Eq. (38)

$$\mu_2 = \mu_w = \mu_w^* + \bar{V}_w \Delta p + RT \ln x_w \tag{47}$$

If box 2 is separated from box 1 and the pressure difference is relieved, then $\Delta p \to 0$ and

$$\mu_w = \mu_w^* + RT \ln x_w \tag{48}$$

which defines the chemical potential of water in an ideal solution. This equation can be generalized to non-ideal solutions and an arbitrary species A, by replacing the mole fraction $x_A$ of species A with its activity $a_A$. Activity is defined to make equation (49) thermodynamically correct for any species in solution whether or not the solution is ideal, i.e.

$$\mu_A = \mu_A^* + RT \ln a_A \tag{49}$$

If we have two fluids (A and B) that mix to form an ideal solution, then the initial Gibbs free energy is

$$G_i = n_A \mu_A^* + n_B \mu_B^* \tag{50}$$

and the final Gibbs free energy is given by

$$G_f = n_A \mu_A + n_B \mu_B \tag{51}$$

where $\mu_A$ and $\mu_B$ are given by Eq. (48), so that the Gibbs free energy of mixing $\Delta G_{\text{mix}} = G_f - G_i$ is given by

$$\Delta G_{\text{mix}} = nRT(x_A \ln x_A + x_B \ln x_B) \tag{52}$$

where $n = n_A + n_B$. Because $\Delta G_{\text{mix}} = -T\Delta S_{\text{mix}}$ and $\Delta H_{\text{mix}} = 0$ for an ideal solution, the total entropy of mixing is given by

$$\Delta S_{\text{mix}} = -nR(x_A \ln x_A + x_B \ln x_B) \tag{53}$$

which is always positive because the mole fractions are less than one, meaning that ideal mixing is always entropically favored. Equation (53) follows from Eq. (41).

## VI. Raoult's law and phase coexistence

### A. Vapor pressure depression

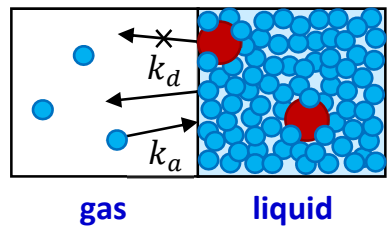

**Fig. 13. Solute-blocking marble game representation of the two-box model of an aqueous solution (liquid) in contact with its vapor (gas). The small blue circles represent water molecules and the larger red circles represent (non-volatile) solutes dissolved in the liquid water. Water molecules can only dissociate (evaporate) from the surface. Solute molecules on the surface block evaporation as indicated by the crossed-out arrow. Water molecules in the gas can associate with (condense on) any portion of the liquid surface, including locations occupied by solute molecules.**

The two-box system of Fig. 13 shows that for a gas-liquid system, dissociation jumps (from box 2 → 1) must occur from the liquid surface and solute molecules block water molecules from reaching a fraction $x_s$ of the surface from the liquid side. As a result, the evaporation rate at the liquid surface is reduced from that for pure water by a factor of $x_w = 1 - x_s$. Modeling this situation with the solute-blocking marble game, results in the FD diagram shown in Fig. 14. The dissociation (evaporation) rate is reduced from that for pure water by a factor of $x_w$ in analogy with the solute-blocking model of osmosis. The association rate (jumps from box $1 \to 2$) is not reduced by the presence of the solute because water molecules can condense on any portion of the liquid surface, including locations occupied by solute molecules.

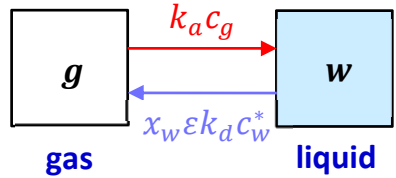

Fig. 14. FD diagram of the two-box gas-water system.

By inspecting the FD diagram, equilibrium occurs when

$$k_a c_g = x_w \varepsilon k_d c_w^* \tag{54}$$

and where from the ideal gas law

$$c_g = \frac{n_g}{V_g} = \frac{p}{RT} \tag{55}$$

In the Gibbs ensemble (constant $(T,P)$) the energy factor is

$$\varepsilon = \exp\left(\frac{-E_b - p\Delta v_{\text{vap}}}{k_B T}\right) \tag{56}$$

where $E_b$ is the binding energy of water molecules in solution and $p\Delta v_{\text{vap}}$ is the work done when a water molecule expands into the gas box at constant pressure. $\Delta v_{\text{vap}} = v_g - v_w$ is the volume change upon vaporization, which can be approximated by $\Delta v_{\text{vap}} \approx v_g$ as $v_g \gg v_w$ at normal temperatures and pressures. Hence,

$$\varepsilon = \exp\left(\frac{-E_b - p v_g}{k_B T}\right) = \frac{1}{e}\exp\left(\frac{-E_b}{k_B T}\right) \tag{57}$$

as for an ideal gas $p v_g = k_B T$. Substituting Eqs. (55) and (57) into Eq. (54) and solving for the pressure results in

$$p = x_w \frac{k_B T}{e v_0} \exp\left(\frac{-E_b}{k_B T}\right) \tag{58}$$

where

$$v_0 = \frac{k_a}{k_d} v_w \tag{59}$$

and $v_w = 1/(N_A c_w^*)$ as before. For pure water, $x_w = 1$ and Eq. (58) reduces to equation (12.17) in Baierlein,[26] which was derived from the semi-classical partition function for a structureless ideal gas and an approximate partition function for an incompressible fluid.

Equation (58) can be rewritten as

$$p = c_0 RT \exp\left(\frac{-E_b}{k_B T}\right) \tag{60}$$

where

$$c_0 = \frac{k_d}{k_a} c_w^* \tag{61}$$

is an empirical parameter.

The Clausius-Clapeyron equation can also be used to derive equation (60) if it is assumed that the enth<u>a</u>lpy of <u>vap</u>orization is given by $\Delta H_{vap} = N_A(E_b + p\Delta v_{vap}) = N_A E_b + RT$ and the liquid binding energy $E_b$ is assumed to be constant, rather than the usual textbook assumption of constant enthalpy $\Delta H_{vap}$.

Equation (58) is Raoult's law

$$p = x_w p^* \tag{62}$$

at any temperature $T$, where

$$p^* = \frac{k_B T}{e v_0} \exp\left(\frac{-E_b}{k_B T}\right) \tag{63}$$

### B. Boiling point elevation

Equation (58) also implicitly predicts boiling point elevation. By setting the pressures of both pure water and the solution to atmospheric pressure, the <u>b</u>oiling <u>t</u>emperature of the solution $T_b$ is related to its water mole fraction by

$$x_w = \frac{T_b^*}{T_b} \exp\left[\frac{E_b}{k_B}\left(\frac{1}{T_b} - \frac{1}{T_b^*}\right)\right] \tag{64}$$

which is approximately linear for small solute mole fractions $x_s$, giving

$$\Delta T_b = K_b b_s \tag{65}$$

where $\Delta T_b = T_b - T_b^*$, $K_b$ is the ebullioscopic constant and $b_s$ is the molality of the solution. Values of $\Delta H_{vap} = 40.68$ kJ/mol and $T_b^* = 373.15$ K predict a value of $K_b = 0.512$ K · kg · mol$^{-1}$ in the physiological range, consistent with literature values. This explains why soup boils at a slightly higher temperature than pure water.

## C. Freezing point depression

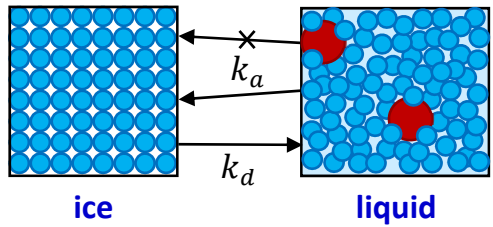

Fig. 15. Solute-blocking marble game representation of the two-box model of an aqueous solution (liquid) in contact with (pure) ice (ice). In reality the two boxes are in direct physical contact, but they have been separated in the diagram to make room for the arrows indicting water molecules associating with the ice (freezing) and dissociating (melting) on the surface of the ice. The small blue circles represent water molecules and the larger red circles represent solutes dissolved in the liquid water. Water molecules can only <u>a</u>ssociate (freeze) at the surface of the ice. Solute molecules on the surface of the liquid block freezing as indicated by the crossed-out arrow. Water molecules in the ice can <u>d</u>issociate from (melt from) any portion of the liquid surface, including locations covered by solute molecules.

Those of us who live in colder climes know that salt will melt ice on the driveway. This freezing point depression is the last colligative property that we will discuss. Fig. 15 is a marble game representation showing the solute-blocking kinetic model of this phenomenon. <u>D</u>issociation jumps now represent the melting of a single water molecule from the surface of the ice into the liquid. The rate of dissociation is not affected by the presence of solute molecules because water can melt on any portion of the ice surface, including locations covered by solute molecules. However, solute molecules at the surface of the ice block liquid water molecules from reaching a fraction $x_s$ of the ice surface and freezing from the liquid side. As a result, the freezing rate at the ice surface is reduced from that for pure liquid water by a factor of $x_w = 1 - x_s$ in analogy with the solute-blocking model of osmosis. Modeling this situation with the solute-blocking model, results in the FD diagram shown in Fig. 16.

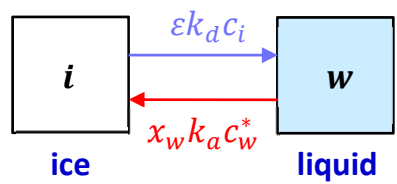

Fig. 16. FD diagram of the two-box ice-water system.

By inspecting the FD diagram (Fig. 16), equilibrium occurs when

$$\varepsilon k_d c_i = x_w k_a c_w^* \tag{66}$$

In the Gibbs ensemble (constant $(T, P)$) the energy factor is

$$\varepsilon = \exp\left(\frac{-E_d + p\Delta v_{\text{fus}}}{k_B T}\right) \tag{67}$$

where $\Delta H_{\text{fus}}/N_A = E_d - p\Delta v_{\text{fus}}$ is the enthalpy of <u>fus</u>ion, per water molecule freezing on the ice, where $E_d$ is the <u>d</u>issociation <u>e</u>nergy for a melting water molecule and $\Delta v_{\text{fus}}$ is the <u>v</u>olume change upon freezing (<u>fus</u>ion). Because the densities of water and ice are similar, $p\Delta v_{\text{fus}} \approx 0$ and $\Delta H_{\text{fus}} \approx N_A E_a$. Hence,

$$\varepsilon = \exp\left(\frac{-\Delta H_{\text{fus}}}{RT}\right) \tag{68}$$

and we find a constant

$$\ln\frac{k_a c_i}{k_d c_w^*} = \frac{\Delta H_{\text{fus}}}{RT_f} + \ln x_w \tag{69}$$

at the equilibrium <u>f</u>reezing <u>t</u>emperature $T_f$.

For pure water, the freezing temperature is $T_f^*$ and $x_w = 1$. Hence, using equation (69) we find that

$$\ln x_w = \frac{\Delta H_{\text{fus}}}{R}\left(\frac{1}{T_f^*} - \frac{1}{T_f}\right) \tag{70}$$

Now $\ln x_w = \ln(1 - x_s) \approx -x_s$ for small solute mole fractions and

$$\frac{1}{T_f^*} - \frac{1}{T_f} = \frac{T_f - T_f^*}{T_f^* T_f} \approx \frac{\Delta T_f}{T_f^{*2}} \tag{71}$$

as $T_f^* T_f \approx T_f^{*2}$ and $\Delta T_f = T_f - T_f^*$. Hence, the freezing point depression is given by

$$\Delta T_f = \frac{-RT_f^{*2}}{\Delta H_{\text{fus}}} x_s \tag{72}$$

which is equation (2.8.30) in Sten-Knudsen[27] that was derived using traditional thermodynamic arguments.

The thermodynamic connection between the colligative properties is well-known.[27] The marble game conceptual framework provides a simple solute-blocking kinetic explanation of all the colligative properties that can in principle be investigated using molecular dynamics simulation techniques in a manner similar to kinetic models of ion channel permeation.[28]

## VII. A pressure gradient in the pore?

According to Kramer and Myers:[10]

> "Osmotic flow is properly described as a bulk or hydrodynamic flow through the pores of the membrane."

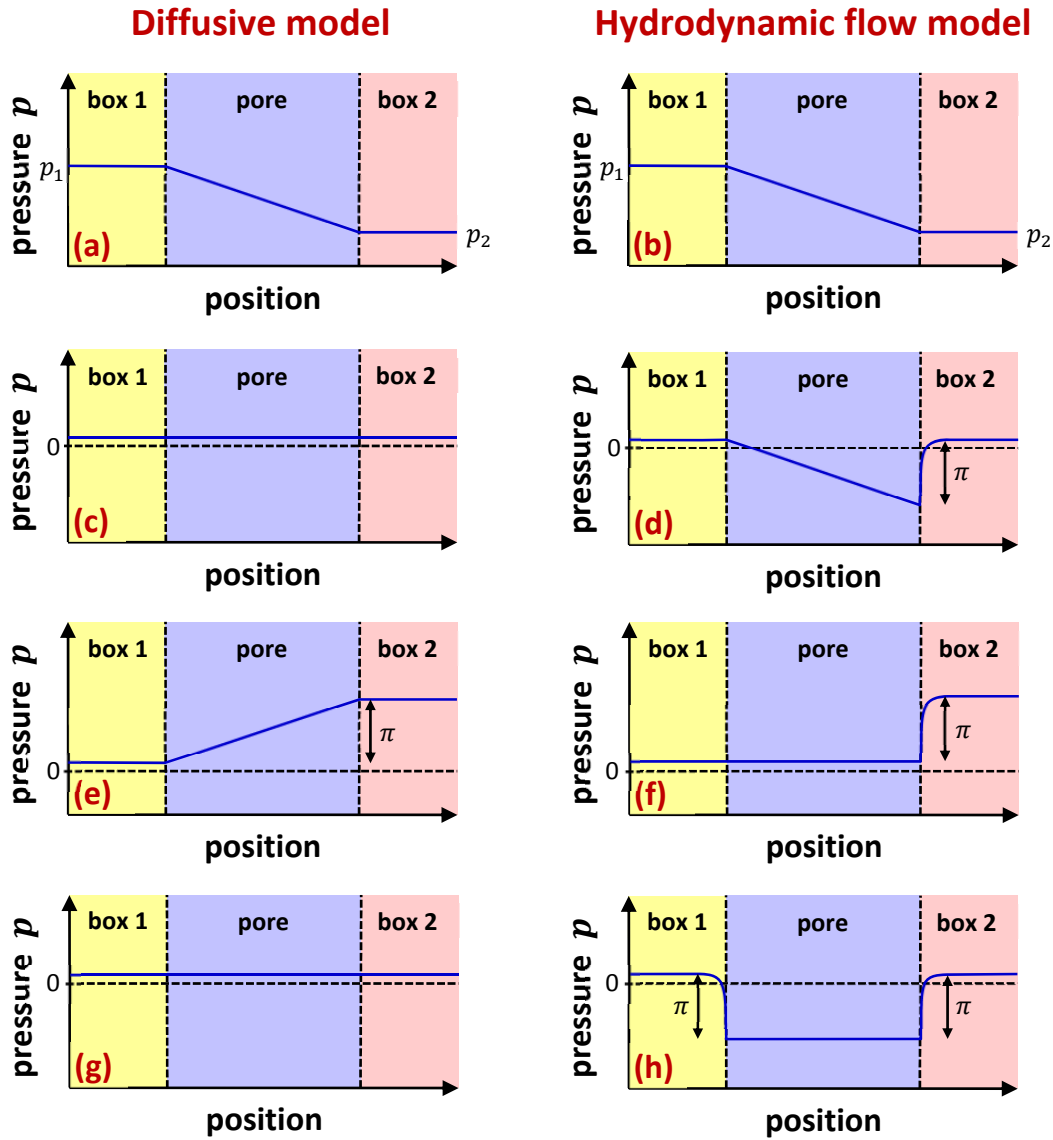

**Fig. 17.** Predictions of the diffusive (marble game) and traditional hydrodynamic flow models of osmosis under various conditions: (a) and (b) pressure driven flow of pure water; (c) and (d) osmotic swelling of a RBC in pure water; (e) and (f) osmotic equilibrium of a rigid plant cell in contact with pure water; and (g) and (h) a RBC in an isotonic solution.

Hence, the traditional hydrodynamic flow model of osmosis advocated by Kramer and Myers includes a pressure gradient within the aquaporin pore whenever the permeation rate is nonzero.[10,11] Within the hydrodynamic flow model of osmosis this is always a real hydrostatic pressure gradient and water in the pore flows like the water in a pipe or blood in a vein or artery. Kramer and Myers' explanation of the origin of this internal pressure gradient is based on an argument that is equivalent to assuming that the pore contains pure (bulk) water and that each end of the pore is in thermodynamic equilibrium with the exterior solution. As a result, there is always a pressure difference between the pore entrance and a solution with non-zero osmotic concentration ($c_s \neq 0$). In the hydrodynamic flow model, that real hydrostatic pressure difference is equal to the osmotic pressure $\pi$ of the solution (Fig. 17(d), (f) and (h)). Figure 17 shows representative examples of the differences between the two models of osmosis. When both boxes contain pure water (Fig. 17(a) and (b)), there is no difference between the two models and there is a gradual pressure gradient along the pore from $p_1$ to $p_2$. This equivalence was not anticipated by those critical of diffusive explanations of osmosis.[10-13]

Whenever box 2 contains solutes ($c_{s_2} \neq 0$), the hydrodynamic flow model predicts a sharp pressure drop of magnitude $\pi$ at the entrance to box 2. Figures 17(c) and (d) show the pressures during the initial osmotic swelling of a RBC in pure water when there is no hydrostatic pressure difference ($\Delta p = 0$). Note that because the marble game model is diffusive, there is no pressure drop required in the pore. However, the hydrodynamic flow model includes a pressure drop of magnitude $\pi$ along the pore which corresponds to a pressure driven flow of pure water within the pore. Also note that the predicted absolute pressure at the box 2 end of the pore is negative for the hydrodynamic flow model, having a value of $p_\text{atm} - \pi = -613$ kPa or about minus six atmospheres. However, the diffusive model never requires negative absolute pressures in the pore and the osmotic pressure difference $\pi$ only appears as a real pressure difference in Fig. 17(e), because the hydrostatic pressure difference $\Delta p = \pi$ for a rigid plant cell in equilibrium with pure water.

## VIII. Tracer counter permeation

The use of tracer-labeled particles in permeation experiments has a long history. In 1955, Hodgkin and Keynes,[17] used radioactive $^{42}$K$^+$ ions to investigate the permeation of potassium ions across the membranes of giant axons from *Sepia officinalis* (common cuttlefish). As a result of comparing tracer counter permeation data with the predictions of the knock-on mechanism they were able to hypothesize that the permeation pathway included two to three single-file K$^+$ ions. This hypothesis was confirmed four decades later when the X-ray structure of potassium ion channels was determined.[29]

Figure 18 shows the same diagram of an aquaporin as Fig. 1 except that now the water in box 1 is tracer-labeled (darker/red) and the water in box 2 is unlabeled (lighter/blue). The water in box 2 is regular H$_2$O, but the water in box 1 is a tracer tagged. This arrangement produces tracer counter permeation (TCP).[19,30] This setup can be achieved experimentally by using deuterated semiheavy water HDO or tritiated water HTO in the solution that a RBC is placed into. The assumption is that the extra neutron or two does not affect the transport properties of the tracer-labeled water. NMR techniques could also be used to tag the molecules.[19]

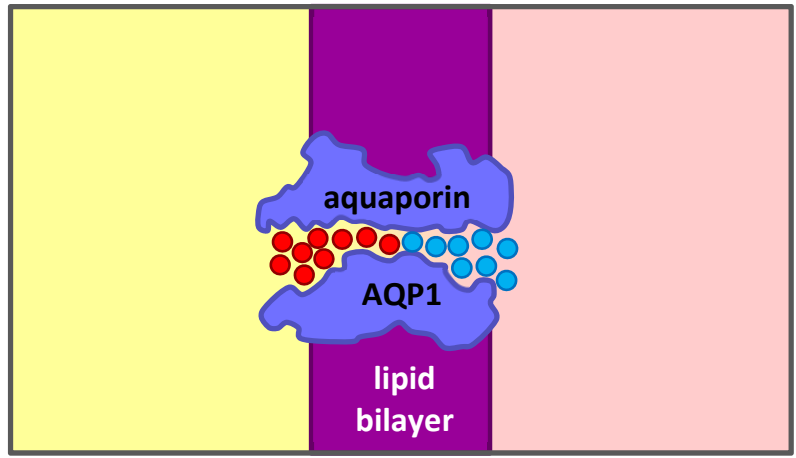

**Fig. 18. Schematic diagram of an AQP1 aquaporin protein (water channel) imbedded in a lipid bilayer membrane with tracer counter-permeation boundary conditions. The solution on the left contains tracer labeled (darker/red) water molecules and the solution on the right contains untagged (lighter/blue) water molecules.**

Figure 19 shows knock-on jump transitions between two possible states of the AQP1 aquaporin selectivity filter (SF) when box 1 contains tracer-labeled water and box 2 contains unlabeled water. Occupancy state 3 corresponds to the arrangement of tracer water molecules shown in Fig. 18 and occurs with probability (occupancy) $\theta_3$ and occupancy state 4 has four tracer-labeled water molecules in the selectivity filter and has occupancy $\theta_4$. A transition from state 3 → 4 occurs when a labeled water molecule enters the SF from the left and an unlabeled molecule exits the right-hand end of the SF. The reverse knock-on jump transition from state 4 → 3 is also shown in Fig. 19. It is assumed that the effective water concentration $c_w$ is the same in both boxes and that there is no pressure difference ($\Delta p = 0$). It is also assumed that only water molecules can be pushed into the SF by thermal fluctuations and cause knock-on jump transitions. The knock-on jump rate $kc_w$ is multiplied by the state occupancy of the originating state because the transition can only occur if the SF starts out in the originating state.

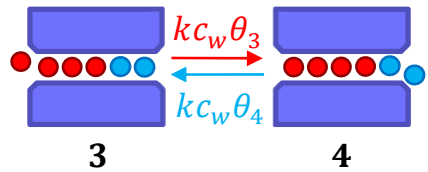

**Fig. 19. Schematic diagram showing knock-on transitions between two occupancy states (3 and 4) of an AQP1 aquaporin selectivity filter with tracer counter-permeation boundary conditions. The filter has $N_{sf} = 5$ single-file water molecules. Occupancy state 3 corresponds to the one shown in Fig. 18. The transition from state 3 → 4 occurs when a tracer-labeled water molecule enters the selectivity filter from box 1 and "knocks on" an unlabeled molecule into the cell. The transition from state 4 to state 3 occurs when an unlabeled water molecule enters the selectivity filter from inside the cell and "knocks on" a tracer-labeled molecule into box 1.**

Figure 20 shows the six possible states of a selectivity filter under TCP boundary conditions that normally contains $N_{sf} = 5$ water molecules. Because the rates of collisions are the same $kc_w$ at both ends of the SF, all of the transitions shown in Fig. 20 are equally likely. Hence, the logical consequence is that all six occupancy states have the same probability. As a result, the steady-state occupancy equation for this system is.

$$\theta_0 = \theta_1 = \theta_2 = \theta_3 = \theta_4 = \theta_5 = \frac{1}{N_{sf}+1} = \frac{1}{6} \qquad (73)$$

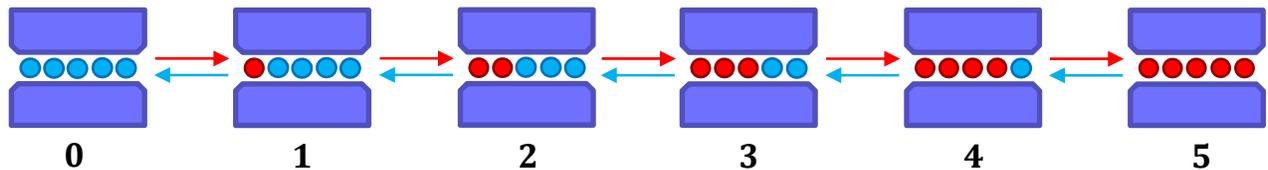

**Fig. 20. Schematic diagram of an AQP1 aquaporin selectivity filter showing the six possible occupancy states of an $N_{sf} = 5$ selectivity filter under tracer counter-permeation (TCP) boundary conditions. All the possible transitions occur with the same probability.**

The transitions shown in Fig. 20 can be viewed as an unbiased one-dimensional random walk of the boundary between the labeled and unlabeled water molecules. Of the states shown in Fig. 20, only state 5 can result in a tracer-labeled water molecule exiting the SF into box 2, as shown in Fig. 21.

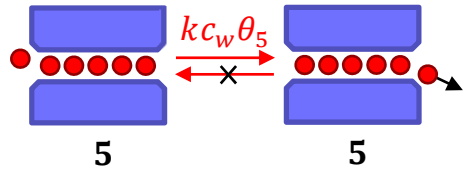

Fig. 21. Schematic diagram of an AQP1 aquaporin selectivity filter showing unidirectional (inward) flux of tracer-labeled water. The forward transition can only happen when the SF is in state 5. The permeant tracer-labeled water molecule is immediately diluted by the unlabeled water molecules on the far side of the aquaporin (in box 2) so that the reverse transition does not happen (as indicated by the crossed-out arrow).

The net effect of the forward knock-on jump transition in Fig. 21 is the jump of a tracer-labeled water molecule from box $1 \to 2$. The reverse transition is assumed to be impossible because the water outside of the aquaporin selectivity filter (in box 2) is well-mixed so that once a tracer-labeled water molecule leaves the selectivity filter, it disappears into the bulk solution of unlabeled water in the RBC and never comes back to the entrance to the selectivity filter.

When the SF is in state 5, the flux through the channel is given by Eq. (9) with $\Delta c_w = -c_w$ for the tracer-labeled water as the tracer concentration in box 2 is zero. However, state 5 only occurs with probability $\theta_5$. Hence, the average tracer flux (the unidirectional flux from box $1 \to 2$) is

$$j_{1\to 2} = \mathcal{P}_f c_w \theta_5 = \frac{\mathcal{P}_f c_w}{N_{sf} + 1} \tag{74}$$

as $\theta_5$ is given by Eq. (73).

The "<u>d</u>iffusive <u>p</u>ermeability" $\mathcal{P}_d$ is defined by analogy with Eq. (9)

$$j_{1\to 2} = \mathcal{P}_d c_w \tag{75}$$

Comparing Eqs. (74) and (75), we find that

$$\frac{\mathcal{P}_f}{\mathcal{P}_d} = N_{sf} + 1 \tag{76}$$

which is the permeability ratio predicted for the knock-on jump mechanism, consistent with the fact that there is a maximum of $N_{sf} + 1$ water molecules in the SF during the knock-on mechanism.[19,31] This permeability ratio was determined experimentally by Mathai *et al.*[22] to be 13.2, which according to Eq. (76) means that there are ~12 single-file water molecules in the aquaporin permeation pathway, which seems consistent with the X-ray structure[16] upon which Figs. 1 and 18 are based. Molecular dynamics simulations are also consistent with this conceptual view.[32]

Equation (76) can be contrasted with the situation where a water molecule can permeate the membrane independently of any others. In that case, the diffusion permeability $\mathcal{P}_d$ under TCP boundary conditions should be the same as the filtration permeability $\mathcal{P}_f$ resulting in the independence relation[17] $\mathcal{P}_f/\mathcal{P}_d = 1$. Permeation examples where the independence relation are expected are the slow permeation of water across a bare lipid bilayer (with no aquaporins) via a dissolve-diffuse-dissolve mechanism, or the permeation of water through an "air membrane" – see Figure 2.7.9 of Feher.[33]

Equation (74) applies when the concentrations are equal and there is no pressure difference. If those restrictions are relaxed, with only tracer-labeled water in box 1 and unlabeled water in box 2, then a kinetic analysis of Fig. 20 with the rates changed to account for the different water mole fractions in boxes 1 and 2, and a nonzero energy factor $\varepsilon$ caused by a nonzero pressure difference $\Delta p$, leads to the prediction that the unidirectional flux (of tracer-labeled water) into box 2 is given by

$$j_{1\to 2} = \frac{\mathcal{P}_f \varepsilon x_{w_1} c_w^* \left(\frac{\varepsilon x_{w_1}}{x_{w_2}}\right)^{N_{sf}}}{\sum_{i=0}^{N_{sf}} \left(\frac{\varepsilon x_{w_1}}{x_{w_2}}\right)^i} \tag{77}$$

and the ratio of the unidirectional fluxes is given by

$$\frac{j_{1\to 2}}{j_{2\to 1}} = \left(\frac{\varepsilon x_{w_1}}{x_{w_2}}\right)^{(N_{sf}+1)} \tag{78}$$

which corresponds to Hodgkin and Keynes' equation (8),[17] with the concentration ratio replaced with the mole fraction ratio and the Boltzmann factor replaced with the energy factor.

## IX. Discussion and conclusion

Despite the fact that the single file nature of osmosis has been known since the late 1950s,[12,34] the view that osmosis is not driven by diffusion is the (current) consensus view of the biophysics[12] and physics[13] communities.[10,11] This is despite the fact that a knock-on model of osmosis was proposed by Lea in 1963.[35] On page 53 of his influential book, Finkelstein discusses the fact that the knock-on mechanism predicts Eq. (76), but he says "This calculation by Lea for tracer *diffusion* is straightforward, but it is unclear what mechanism for osmotic *flow* he has in mind that allows him to thereby conclude that $P_f/P_{d_w} = N + 1$, a result almost identical to eq. (4-17)."[12] (emphasis added). As a result, it seems clear that the knock-on model was rejected as a model of osmosis because the consensus view was that osmosis must be the hydrodynamic *flow* of water through a narrow pore driven by a pressure gradient (page 19),[12] whereas the knock-on mechanism models osmosis as a *diffusive* process. Figure 17 summarizes differences in the pressure profiles for the diffusive model presented here and the traditional hydrodynamic flow model. The predictions of these two competing models can (in principle) be tested using molecular dynamics simulations.[32] The predictions of the knock-on model under TCP boundary conditions have already been confirmed by experiment and molecular dynamics simulations.[22,32]

A central theme of the marble game approach to molecular modeling is "thermodynamics from kinetics".[8] That philosophy has been extended here to address the long-running controversy surrounding osmosis and to explaining the colligative properties of dilute solutions, the entropy of mixing, free energies and the central role of the chemical potential in transport phenomena. These are topics that should be considered for inclusion in the redesign of introductory physics courses for the life sciences.[4]

## Acknowledgments

I wish to thank Eileen Clark, Jaqui Lynch, Niina Ronkainen and the boys for helpful comments on an earlier draft of the manuscript. Support from the National Institutes of Health (Fellowship GM20584) and National Science Foundation (Grant No. 0836833) is gratefully acknowledged.